\documentclass[prb,floats,showpacs,aps,twocolumn,floatfix,amsmath]{revtex4}
\usepackage{graphicx}
\begin{document}
\title{Helical scattering and valleytronics in bilayer graphene}

\author{Henning Schomerus}
\affiliation{Department of Physics, Lancaster University,
Lancaster,  LA1 4YB, United Kingdom}


\date{\today}

\begin{abstract}
We describe an angularly asymmetric interface-scattering mechanism which
allows to spatially separate the electrons in the two low-energy valleys of
bilayer graphene. The effect occurs at electrostatically defined interfaces
separating regions of different pseudospin polarization, and is associated
with the helical winding of the pseudospin vector across the interface, which
breaks the reflection symmetry in each valley. Electrons are transmitted with
a preferred direction of up to $60^\circ$ over a large energetic range in one
of the valleys, and down to $-60^\circ$ in the other. In a Y-junction
geometry, this can be used to create and detect valley polarization.
\end{abstract}

\maketitle

\section{Introduction}

Research on low-dimensional systems has received a significant stimulus by
the recent isolation of atomically thin sheets of graphite, known as graphene.\cite{discovery} Graphene's attractive electronic properties
can be traced back to a unique low-energy band structure (corresponding to
massless Dirac particles in pristine monolayers, which become massive in
bilayers), whose key features are dictated by the existence of an intrinsic
orbital degree of freedom, known as \emph{pseudospin}---a choice of two $\pi$
orbitals, presented by the two chemically identical carbon atoms in the  unit
cell of the graphene honeycomb lattice.\cite{review}
The celebrated phenomenon of chirality ties the pseudospin to the momentum,
which results in a Berry phase of $\pi$ in monolayers ($2\pi$ in bilayers) whose consequences are observed in the
quantum Hall effect \cite{qhe1,qhe2,qhe3} and in the chiral Klein tunneling across np junctions.\cite{Katsnelson,Cheianov}

A descended, remarkably robust electronic feature---shared by graphene
monolayers, bilayers, and multilayers of various stacking order---is the
splitting of the low-energy band structure into two energetically degenerate
but orbitally inequivalent \emph{valleys} (of different {\em isospin}),
situated around the corners of the Brillouin zone (the K and K$'$ points). In
this work we demonstrate that the electrons belonging to the valleys can be
separated and detected by a variant of the Klein tunneling set-up, which
utilizes a certain type of electrostatically defined interfaces in bilayer
graphene. The interfaces in question separate regions of opposite pseudospin
polarization, and are obtained by combining pairs of oppositely charged top-
and back gates as illustrated in Fig.\ \ref{fig1}(a). In the past, such
interfaces have attracted attention because they guide valley-dependent
states \cite{blanter} that propagate along the interface; for transport in
the perpendicular direction, they realize a pseudospin variant of a spin
valve,\cite{sanjose2009} which allows electronic confinement.\cite{peeters} By resolving the transport across the interface
angularly in each valley, we find that the direction of perfect transmission  is
skewed away from normal incidence, see Fig.\
\ref{fig1}(c)-(h). The broken reflection symmetry is intimately related to a
hidden helical structure
---the winding of the pseudospin vector as one crosses the interface, which is opposite in the two valleys
as illustrated in Fig.\ \ref{fig1}(b).
The ensuing helical scattering favors transmission at opposite angles of
incidence, and provides a means to separate electrons in the valleys by
directing them to different leads.

From a practical perspective, efforts to create an imbalance in the
population of the valleys are motivated by the prospects of electronic
applications that do not directly rely on the charge of the carriers.\cite{valleytronics}
Dubbed valleytronics, this scheme aims to implement
analogies to the rapidly maturing field of spintronics,\cite{spintronics}
with the two valleys substituting the two states of physical electronic spin.
The original proposal to create and detect valley polarization relies on edge
effects;\cite{valleytronics} alternative proposals employ broken sublattice
symmetry combined with in-plane electric fields,\cite{wang} or drastic
doping.\cite{cortijo} Besides the practical obstacles to implement these
routes, these proposals face the severe problem of strong intervalley
scattering occuring at pn junctions in monolayer graphene.\cite{akhmerov} In
this context,  the valley-dependent helical scattering offers two key
advantages: The required interfaces can be implemented and controlled by
ordinary gates, and intervalley scattering is negligible. Towards the end of this paper, we evaluate the feasibility of this route to valleytronics by
numerical results for wide (but finite-sized) Y-junction devices.

The paper is organized as follows:
Section \ref{sec2} presents analytical results for scattering at pseudospin interfaces in
bilayer graphene.
Section \ref{sec3} presents numerical results for valleytronics applications.
Section \ref{sec4} contains conclusions. Appendices provide background for the analytical and numerical methods.

\section{\label{sec2}Helical scattering}

\begin{figure}
\includegraphics[width=.95\columnwidth]{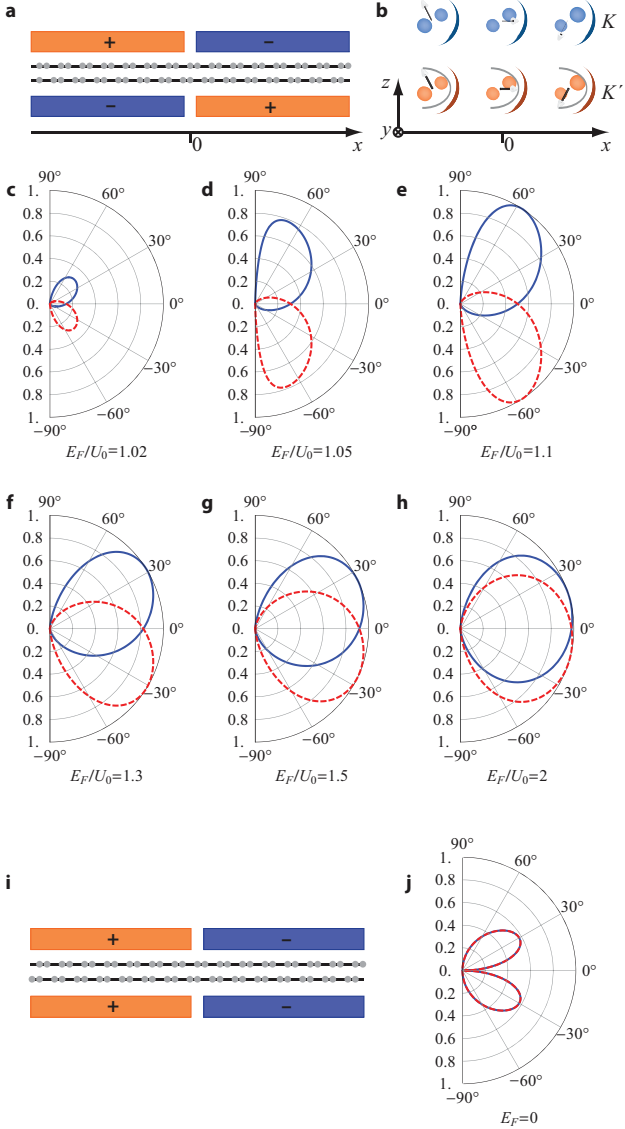}
\caption{\label{fig1} (Color online).
Helical scattering of electrons in
bilayer graphene at an interface between regions of different pseudospin
polarization. Panel (a) sketches in a side-view how such an interface is obtained by sandwiching
a bilayer graphene flake between positively and negatively charged gates (orange and blue,
respectively; not drawn to scale). Panel (b) shows the rotation of the pseudospin vector in the valleys around the K and K$'$ points.
In the polar plots of panels
(c)-(h), solid (blue) lines show the angle-resolved transmission
probability $T(\varphi)$ in the K valley, while dashed (red)
lines show this probability in the  K$'$ valley. Each panel
pertains to a different Fermi energy  $E_F$, ranging from values just above
the gap  where the transmission in each valley is angularly asymmetric but
small  [panels (c) and (d)], over a regime where angular asymmetric
scattering occurs around a transparent direction [panels (e)-(g)] to
the regime far above the gap, where one approaches valley-independent,
angularly symmetric scattering [panel (h)].
Panels (i) and (j) contrast these finding to the case of Klein tunneling at an np junction,\cite{Katsnelson,Cheianov}
which does not distinguish the valleys. }
\end{figure}

We start
our considerations with the scattering at pseudospin interfaces in bulk
graphene, which allows a fully analytical treatment that reveals the
connection of valley-dependent scattering and helical winding.

In bilayer graphene, pseudospin refers to the probability amplitude to find
an electron on the top layer (spin up, amplitude $\psi_\uparrow({\bf r})$) or on
the bottom layer (spin down, amplitude $\psi_\downarrow({\bf r})$). Pseudospin
polarization can be created by oppositely charged top- and back gates, which
induce an electrostatic potential $U_\uparrow=-U_\downarrow\equiv U(x)$ in
the two layers. Patterning of the gates allows to spatially vary the
potential. The relevant low-energy electronic transport in this landscape can
be described by a two-component Hamiltonian \cite{bilayer}
\begin{equation}
H=\left(
                     \begin{array}{cc}
                       U(x) & -(\hat p_x-i\xi \hat p_y)^2/2m \\
                       - (\hat p_x+i\xi\hat p_y)^2/2m & -U(x) \\
                     \end{array}
\right),\label{eq:h}
\end{equation}
which acts on the spinor wave function $\psi({\bf r})=(\psi_{\uparrow}({\bf r}),
\psi_{\downarrow}({\bf r}))^T$.  Here $m$ is the effective mass, and $\hat p_x$,
$\hat p_y$ are the two components of the momentum operator in the graphene
plane. The valley index $\xi$ takes the values $\xi=1$ in the K valley, and
$\xi=-1$ in the K$'$ valley.

A sharp pseudospin interface along the line $x=0$ is described by a step-like
profile of the electrostatic potential, with $U(x)=-U_0$ for $x<0$ and
$U(x)=U_0$ for $x>0$. In the regions of constant potential, the electronic
bilayer spectrum is gapped, so that electrons can only move freely for Fermi
energies $E_F$ fulfilling $|E_F|>\Delta/2 = |U_0|$. (At energies within the
gap, there is a pair of states which propagate along the interface;\cite{blanter} these states do not contribute to the transport across the
interface. In the following, we will assume $E_F>U_0>0$.) In each valley, the
angle-resolved probability $T(\varphi)=|t(\varphi)|^2$ of electrons
approaching the interface from the left to be transmitted to the right can be
calculated in a standard wave-matching procedure (see Appendix \ref{app1}). The
transmission amplitude is
\begin{equation}
t=2\frac{i p_y^2(1-\beta^2){\rm Im}\,\alpha+ip_x q[{\rm
Re}\,\alpha(1+\beta^2)-\beta(1+|\alpha|^2)]}
{p_y^2(1-\beta^2)(1-\alpha^{*2})+ip_xq[(\alpha^*-\beta)^2+(1-\alpha^*\beta)^2]}
,\label{eq:result}
\end{equation}
with $p_y=p_0\sin\varphi$, $p_x=p_0\cos\varphi$,  $q=p_0\sqrt{1+\cos^2\varphi}$,  $\alpha=-\frac{(p_x+i\xi p_y)^2}{2m(E_F-U_0)}$,
$\beta=\frac{(q-\xi p_y)^2}{2m(E_F-U_0)}$, and
$p_0=\sqrt{2m}(E_F^2-U_0^2)^{1/4}$.
The resulting angle-resolved probability only depends on $E_F/U_0$, and is plotted in Figure
\ref{fig1}(c)-(h) as a sequence of polar plots.

Just above the gap ($E_F=1.02\,U_0$), the transmission probability in both
valleys is small for any angle of incidence (this is the essence of the
pseudospin valve effect \cite{sanjose2009}). As the Fermi energy increases,
the transmission probability increases,
 but in an angularly asymmetrical fashion: over a
large range of energies, electrons in the K valley have large transmission
probability for incident angles in the range of $30^\circ$ to $60^\circ$,
where the interface is almost transparent. The transmission characteristics
in the K$'$ valley are similarly skewed, but into the opposite direction. An
approximately symmetric, valley-independent transmission probability only
emerges  for energies of the order of $E_F=2U_0$ and beyond.
These observations can be quantified analytically by considering the
direction of unit transmission probability, which exists for values $E_F\geq
\sqrt{9/8}U_0\approx 1.06\,U_0$, and then is given by the momentum vector
\begin{equation}
{\bf p}_{\rm 0}=\left(
          \begin{array}{cc}
            \sqrt{3m\sqrt{E_F^2-U_0^2}-m E_F} \\
            \displaystyle
                   \frac{\xi}{2}\left(\sqrt{2m(E_F+U_0)}-\sqrt{2m(E_F-U_0)}\right)
          \end{array}
        \right).
        \label{eq:transp}
\end{equation}
The component parallel to the interface is finite, and has an opposite sign
in the two valleys; it only becomes negligible when $E_F\gg U_0$.

It is natural to inquire whether there is a fundamental mechanism behind this
valley-dependent transmission pattern. Here we offer two
observations---firstly, the helical winding of the pseudospin vector for
propagating modes as one crosses the interface, which manifests the broken
reflection symmetry in each valley; secondly, a condition for the pseudospin vector of
evanescent modes, which identifies the transparent direction $\textbf{p}_0$.

The pseudospin vector $\vec P=\langle \vec{\boldsymbol\sigma}\rangle$ is a unit vector
composed of the expectation values of the Pauli spin matrices. For a constant
potential $U$, a propagating wave with momentum $\bf p$ carries pseudospin
\begin{equation}
\vec P_{\rm prop}=\frac{p_y^2-p_x^2}{2mE_F}\hat {\bf i} -\frac{\xi p_x p_y}{m E_F}
\hat {\bf j}+\frac{U}{E_F}\hat {\bf k}.
\label{pseudospinvector}
\end{equation}
 The out-of-plane component is a measure of
the pseudospin polarization, while the projection onto the graphene plane is
 linked to the propagation direction (this is the celebrated chirality of charge
carriers in graphene \cite{review}); notably, the sign of the $y$ component
depends on the valley index $\xi$. Far to the left and to the right of the
interface, the pseudospin vector has a finite but opposite out-of-plane
component.  As $U$ changes from $-U_0$ to $U_0$ the pseudospin vector
describes an arc on the unit sphere, which has opposite orientation in the
two valleys (see Fig.\ \ref{fig1}(b)), and thereby explicitly breaks the reflection symmetry in a fixed
valley.  The asymmetric transmission pattern is therefore connected to a
hidden helical structure in each of the two valleys.

Our second observation involves the evanescent modes, which carry pseudospin
close to the interface, and whose amplitudes are related to the propagating
modes by the wave matching conditions. At the transparency condition
(\ref{eq:transp}), the evanescent modes on both sides of the interface have
the same amplitude and phase on both layers, $\psi_{\rm ev}\propto(1,1)^T$.
Their pseudospin vector ${\vec P}_{\rm ev}=\hat{\bf i}$ therefore lies in the
graphene plane, and points into the direction perpendicular to the interface.
While this condition is the same in both valleys, the properties of the
evanescent modes differ; as a result, they meet the condition for the finite,
valley-dependent value of $p_y=p_{0y}$ specified in Eq.\ (\ref{eq:transp}),
which again explicitly breaks the reflection symmetry in each valley.

\section{\label{sec3}Valleytronics applications}
\begin{figure}
\includegraphics[width=.85\columnwidth]{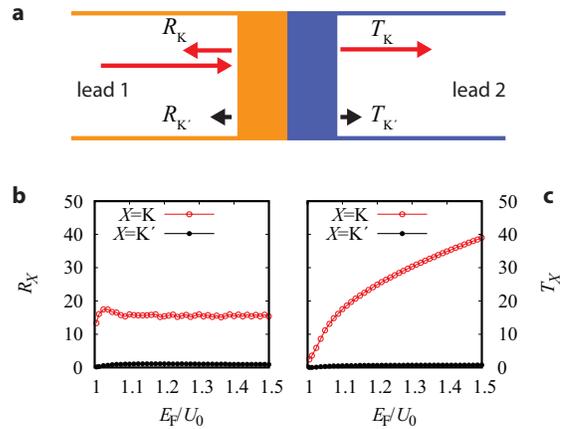}
\caption{\label{fig2} (Color online). Absence of intervalley scattering.
Panel (a): Top-view sketch of a (zigzag) graphene nanoribbon geometry with a pseudospin interface, and definitions of reflection and transmission
coefficients (summed over all channels) when a K-polarized current is injected into lead 1.
Panels (b) and (c): Numerical results for the valley-resolved reflection and transmission coefficients (nanoribbon width $W\simeq 150 \hbar/\sqrt{2 m |U_0|}$).}
\end{figure}

\begin{figure}
\includegraphics[width=\columnwidth]{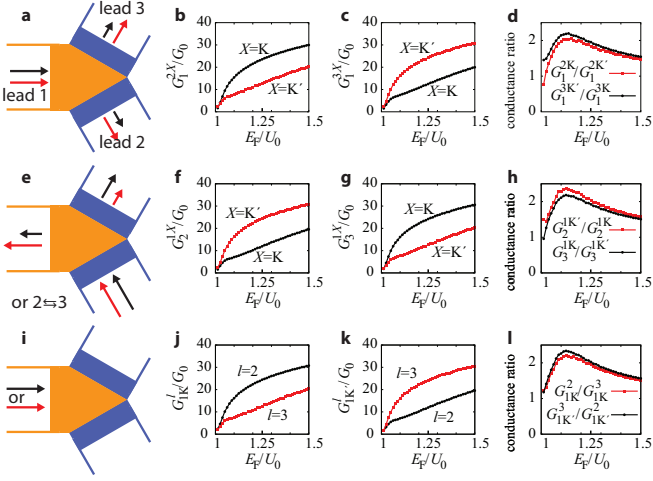}
\caption{\label{fig3} (Color online).
Valleytronics applications.
Panel (a): Top-view sketch of a valley
splitter, consisting of a Y-junction where electrons are injected through
lead 1, and collected in leads 2 and 3. In this set-up, the contact region
has the same gate polarity as lead 1, and the incoming current in  this lead
is valley-unpolarized. Panels (b) and (c): Valley-resolved conductance of
electrons in leads 2 and 3. Panel (d): Ratio of the larger to the smaller
conductance in these two leads. Analogously, panels (e)-(h) demonstrate that
valley polarization can be created in lead 1 by injecting unpolarized
currents into lead 2 or 3. Panels (i)-(l) show how the total current in leads
2 and 3 can be used the detect the valley polarization of an incoming current
in lead 1.}
\end{figure}

We now turn to the prospects to employ helical scattering for valleytronics
applications. In view of previous proposals in monolayer graphene,
\cite{valleytronics,wang,cortijo} a key question is whether pseudospin
interfaces induce intervalley scattering in zigzag nanoribbons, as this is
the orientation in which the valleys are conserved away from the interface.
This question goes beyond the continuum description,  \cite{akhmerov} and is
numerically addressed in  Figure \ref{fig2}. The top panel shows the
nanoribbon geometry. To quantify the intervalley scattering, consider the
current $I_{1K}^{lX}$ in lead $l=1,2$ and valley $X$=K, K$'$ in response to a
bias voltage $V$ applied to lead 1, where the electrons are polarized in
valley K (below we describe how such valley polarization can be created; lead
2 is grounded). For small bias we can define valley- and lead-resolved
conductances $G_{1K}^{lX}=I_{1K}^{lX}/V$, which can be evaluated using
standard quantum transport algorithms (see Appendix \ref{app2}). The bottom panels show
valley-resolved transmission coefficients $T_X=G_{2X}^{1K}/G_0$ (where $G_0$
is the conductance quantum), as well as reflection coefficients $R_X$,
defined analogously based on the reflected electron stream, for a wide nanoribbon
of width $W\simeq 150 \hbar/\sqrt{2 m |U_0|}$ (for these wide ribbons we do
not find any difference between the zigzag and antizigzag edge configurations
\cite{akhmerov}). The results show that intervalley scattering is negligible
both in reflection as well as in transmission. Encouragingly, the
transmission coefficient exceeds the reflection coefficient once the regime
of helical scattering is entered. The results also reveal that for the
assumed sharp interface the reflection coefficient is almost
energy-independent (softening the interface further reduces the reflection
and intervalley scattering).

In a suitable geometry, the helical scattering should therefore enable to
spatially separate the electrons in the two valleys. To evaluate the
feasibility of such a bulk valley filter we consider a Y-junction geometry
with three leads, shown in Figure \ref{fig3}(a), where a pseudospin interface
is placed at the entrance of leads 2 and 3. This is done with the intent that
electrons injected into lead 1 are preferably transmitted into lead 2 when
they are in the K valley, and into lead 3 when they are in the K$'$ valley.
To quantify the anticipated valley polarization, we now consider the current
$I_1^{lX}$ in lead $l=2,3$ and valley $X$=K, K$'$ in response to a bias
voltage $V$ applied to lead 1 (the injected current from this lead is now
valley-unpolarized; leads 2 and 3 are both grounded), and the associated
conductances $G_1^{lX}=I_1^{lX}/V$. Figures \ref{fig3}(b)-(d) show the energy
dependence of these conductances, along with the ratio of the larger over the
smaller conductance in each lead. The results demonstrate that a finite
valley polarization can indeed be obtained for conditions where the
conductance itself is also large ($E_F\approx 1.1 U_0$, where the preferred
transmission directions are $\approx \pm 60^\circ$). By injecting unpolarized
currents into leads 2 or 3, the Y-junction geometry can also be used to
create valley polarization in lead 1; this is shown in panels (e)-(h). Panels
(i)-(l) show how valley polarization in lead 1 is detected by the total
current in leads 2 and 3, which is quantified by conductances $G_{1X}^l$ (now
$X$ stands for the valley polarization of the incoming current). A K
valley-polarized current injected into lead 1 will preferably divert to lead
2 (panel j), while a K$'$  valley-polarized current will preferably divert to
lead 3 (panel k).

\begin{figure}
\includegraphics[width=\columnwidth]{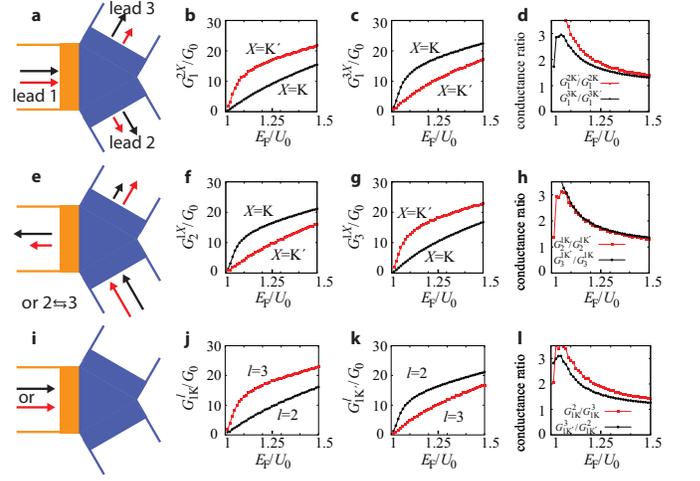}
\caption{\label{fig4} Alternative design of the valley splitters, polarizers
and detectors of Fig.\ \ref{fig3}, where the contact region
has the same gate polarity as leads 2 and 3. Qualitatively, the results for
valleys K and K$'$ are now interchanged; however, the influence of edge
states between leads 2 and 3 result in a notable lead-asymmetry of the
conductance ratios.}
\end{figure}

Figure \ref{fig4} shows an alternative design of the Y-junction, where now the contact region has the same gate
polarity as leads 2 and 3. This produces larger conductance ratios for valley
polarization and detection, but suffers from larger asymmetries which
originate from edge states that connect leads 2 and 3. Symmetry upon
interchanging the lead index is restored when one simultaneously inverts the
sign of $U_0$ and $E_F$.

\section{\label{sec4}Conclusions}

In summary, we demonstrated that electrostatically defined pseudospin interfaces in bilayer graphene result
in angularly asymmetric, valley-dependent scattering. Above a certain energy threshold, the transmission
becomes perfect into a particular direction, which is opposite in the two valleys. This direction only turns towards normal to the interface when energies are very large.
The broken reflection
symmetry in each valley can be made manifest by considering the helical
winding of the pseudospin vector across the interface.

From a conceptual
point of view, our findings enforce the role of interfaces to unravel
fundamental properties of charge carriers in graphene, thereby complementing
the earlier instances of np junctions (which allow to realize Klein tunneling
\cite{Katsnelson,Cheianov,Young} and Veselago lensing \cite{chei2}), flake boundaries (which can support edge states
\cite{edgestates}), and regions of large charge carrier density (which induce
surface states that support tunneling over large distances \cite{evmodes}).
From a more practical perspective, the helical scattering opens an avenue to
manipulate the valley degree of freedom in bulk graphene.

The author gratefully acknowledges discussions
with Elsa Prada, Pablo San-Jose, and Edward McCann.

\appendix

\section{\label{app1}Wave matching}

The transport problem across the pseudospin interface can be solved by matching the propagating and evanescent modes
to the left and right of the interface, which are obtained from the
Schr{\"o}dinger equation $H\psi=E\psi$ with Hamiltonian $H$ given in Eq.\
(\ref{eq:h}).

The scattering at the interface conserves energy $E$ and the component
$p_y$ of the momentum parallel to the interface. In the region $x<0$, where
$U=-U_0$, an incident mode with fixed $p_y$ and $E$ has longitudinal momentum
$p_x=\sqrt{2m\sqrt{E^2-U_0^2}-p_y^2}\equiv p$, and is described by a wave
function
\begin{equation}\psi_{\rm in}=\exp(ip x+ip_yy)
\left(
                          \begin{array}{c}
                            1 \\
                            \alpha   \\
                          \end{array}
                        \right)
,
\end{equation}
where $\alpha=-\frac{(p+i\xi p_y)^2}{2m(E-U_0)}$. The reflected wave
\begin{equation}\psi_{\rm ref}=\exp(-ip x+ip_yy) \left(
                          \begin{array}{c}
                            1 \\
                            \alpha^*   \\
                          \end{array}
                        \right)
\end{equation}
has longitudinal momentum $p_x=-p$. There is also an evanescent mode
\begin{equation}\psi_{\rm left}=\exp(q x+ip_y y) \left(
                          \begin{array}{c}
                            1 \\
                            \beta   \\
                          \end{array}
                        \right)
,
\end{equation}
which decays (to the left, away from the interface) with imaginary momentum $p_x=-iq$,
$q=\sqrt{2m \sqrt{E^2-U_0^2}+p_y^2}$, and $\beta=\frac{(q-\xi
p_y)^2}{2m(E-U_0)}$.

In the region $x>0$, where $U=U_0$, the transmitted wave
\begin{equation}\psi_{\rm trans}=\exp(ip x+ip_yy) \left(
                          \begin{array}{c}
                            \alpha^* \\
                            1   \\
                          \end{array}
                        \right)
\end{equation}
has longitudinal momentum $p_x=p$, and the decaying evanescent mode is given
by $p_x=iq$,
\begin{equation}\psi_{\rm right}=\exp(-q x+ip_yy) \left(
                          \begin{array}{c}
                            \beta \\
                            1   \\
                          \end{array}
                        \right)
.
\end{equation}

The wave matching conditions at the interface $x=0$ between the regions are
continuity,
\begin{equation}
\left( \begin{array}{c} 1 \\ \alpha   \\ \end{array}\right)
+r
\left( \begin{array}{c} 1 \\ \alpha^*   \\ \end{array}\right)
+c
\left( \begin{array}{c} 1 \\ \beta   \\ \end{array}\right)
=t
\left( \begin{array}{c}  \alpha^* \\1  \\ \end{array}\right)
+d\left( \begin{array}{c} \beta \\ 1  \\ \end{array}\right)
,
\end{equation}
and continuity of the first derivative,
\begin{equation}
p\left( \begin{array}{c} 1 \\ \alpha   \\ \end{array}\right)
-p r
\left( \begin{array}{c} 1 \\ \alpha^*   \\ \end{array}\right)
-i qc
\left( \begin{array}{c} 1 \\ \beta   \\ \end{array}\right)
=p t
\left( \begin{array}{c}  \alpha^* \\1  \\ \end{array}\right)
+i q d\left( \begin{array}{c} \beta \\ 1  \\ \end{array}\right)
.
\end{equation}
Together, they form an inhomogeneous system of four linear  equations which
uniquely determines the four coefficients $r$, $t$, $c$, and $d$, where in
particular
\begin{equation}
t=\frac{2 i p_y^2(1-\beta^2){\rm Im}\,\alpha+2ipq[{\rm
Re}\,\alpha(1+\beta^2)-\beta(1+|\alpha|^2)]}
{p_y^2(1-\beta^2)(1-\alpha^{*2})+ipq[(\alpha^*-\beta)^2+(1-\alpha^*\beta)^2]}
.\label{eq:result}
\end{equation}
The transmission probability follows from $T=|t|^2$ (the corresponding
reflection probability is $R=|r|^2=1-T$). (In the main text, we renamed $E\to E_F$, and $p\to p_x$.)
Condition (\ref{eq:transp}), which
singles out the propagation direction ${\bf p}_0$ with unit transmission
$T=1$, coincides with the condition $\beta=1$. The transmission amplitude
then takes the value $t=(\alpha-1)/(1-\alpha^*)$.

\section{\label{app2}Numerical Method}

\begin{figure}
\includegraphics[width=.9\columnwidth]{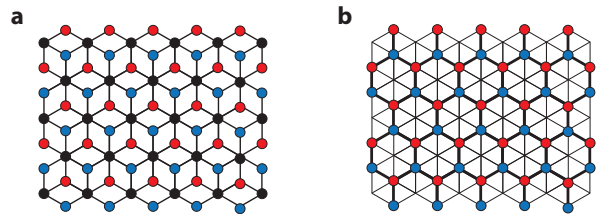}
\caption{\label{figS1}
(a) Tight-binding model of a flake of bilayer graphene, consisting of a
top and a bottom layer where each layer accommodates a honeycomb lattice of
carbon atoms. Black dots denote dimer sites, where carbon atoms of both
layers lie on top of each other and are coupled with inter-layer coupling
constant $\gamma_1$. Light (red and blue) dots denote  sites of isolated carbon
atoms (either on the top or the bottom layer). Lines denote intra-layer nearest-neighbor
hopping with coupling constant $\gamma_0$. (b) Effective model for the two low-energy bands, obtained after
decimation (algebraic elimination) of the dimer sites. Thick lines denote
coupling with strength $2\gamma'$, while thin lines denote coupling with
strength $\gamma'$, where $\gamma'=\gamma_0^2/\gamma_1$. }
\end{figure}

We base our numerical calculations 
on a tight-binding model which describes the two low-energy bands of bilayer
graphene. This model is obtained from the physical bilayer graphene lattice
of carbon atoms, shown in Fig.\ \ref{figS1}(a), where the $\pi$ orbitals of
neighboring atoms
 in the same layer are coupled by an intra-layer  hopping
energy $\gamma_0$, while orbitals of atoms that lie on top of each other
(\emph{dimer sites}) are coupled by the inter-layer hopping energy
$\gamma_1$. At low energies, the latter atoms are only visited in virtual
transitions, and can be eliminated using decimation. This amounts to solving
the Schr{\"o}dinger equation for the amplitudes on the dimer sites, and
inserting the result back into the remaining equations of the non-dimer
sites. The (generally energy-dependent) self energies and mediated hoppings
can be set to their values at the Dirac point $E=0$. The resulting
two-component tight-binding model is shown in Fig.\ \ref{figS1}(b). Nearest
neighbors are now coupled with strength $2\gamma'$, where
$\gamma'=\gamma_0^2/\gamma_1$ is an effective hopping energy. Coupling to
next-nearest neighbors is still constrained to atoms of different
sublattices, and carries strength $\gamma'$. This model allows to efficiently
enter the regime of many propagating modes in the leads without interfering
with the split bands.

The valley-resolved conductances are determined from the relation
\[
G^{lX}_1=G_0\!\!\!\!\!\!\!\!\!\!\!\sum_{{n\in \mbox{\scriptsize lead $l$, valley $X$}}\atop{m\in \mbox{\scriptsize lead
1}}}|S_{nm}|^2.
\]
 Here $G_0=2e^2/h$ is the conductance quantum, where the
factor 2 arises from the degeneracy of the physical spin, and  $S_{nm}$ are
components of the scattering matrix, which we calculate by a standard
numerical decimation method. \cite{decimation} The conductances $G_{l}^{1X}$ and  $G_{1X}^l$ are defined and obtained analogously.
The valleys are resolved by
using the Bloch wave propagation factor $\lambda=\exp(i\sqrt{3}\,a k_x)$ of
modes in leads aligned along the valley-preserving zigzag orientation (here
$a$ is the intra-layer distance of neighboring carbon atoms). In the K
valley, ${\rm Im}\, \lambda<0$, while in the K$'$ valley ${\rm Im}\,
\lambda>0$.

\end{document}